\documentclass[prd,11pt,floatfix,superscriptaddress,showkeys]{revtex4}

\usepackage{latexsym,amsbsy}
\usepackage[dvips]{graphicx}
\DeclareGraphicsExtensions{.eps, .jpg}
\RequirePackage[dvips,colorlinks,linkcolor=blue,citecolor=blue]{hyperref}
\begin{document}
\setcounter{page}{1}
\thispagestyle{empty}

\title{Modified Equations of State for Dark Energy\linebreak  and Observational
Limitations}

\author{German S. Sharov$^{1,2}$* 
  and Vasily E. Myachin 
}  

\affiliation{Department of Mathematics, Tver State University, Sadovyi per. 35, Tver,
Russia\\
$^{2}$ International Laboratory for Theoretical Cosmology, Tomsk State University\\ of
Control Systems and Radioelectronics (TUSUR), 634050 Tomsk, Russia }

\email{Sharov.GS@tversu.ru}

\begin{abstract}  
 Cosmological models with variable and modified equations of state for dark
energy are confronted with observational data, including Type Ia supernovae, Hubble
parameter data $H(z)$ from different sources, and observational manifestations of cosmic
microwave background radiation (CMB). We consider scenarios generalizing the
$\Lambda$CDM, $w$CDM, and Chevallier--Polarski--Linder (CPL) models with nonzero
curvature and compare their predictions. The most successful model with the dark energy
equation of state $w= w_0+ w_1(1-a^2)/2$ was studied in detail.
 These models are interesting in possibly alleviating the Hubble
constant $H_0$ tension, but they achieved a modest success in this direction with the
considered observational data.
 \end{abstract}


 \maketitle

\section{Introduction}\label{Intr}

In modern cosmology based on recent observational data, our Universe includes dominating
fractions of dark energy and dark matter, whereas all kinds of visible matter fill about
4\% in total energy balance nowadays. The~latest estimations of Planck collaboration~\cite{Planck15,Planck18} predict about 70\% fraction of  dark energy,  if~we apply the
standard $\Lambda$CDM model, where dark energy may be represented as the cosmological
constant $\Lambda$ or as a matter with density $\rho_x$ and pressure $p_x=-\rho_x$.
Almost all remaining part of matter in this model is cold dark matter with close to zero
pressure. Because~of the last property, it is convenient to consider cold dark matter
together with visible baryonic matter, where the~unified density is $\rho_m=\rho_c+\rho_b$.
One should also add the radiation component including relativistic species (neutrinos)
with $p_r=\frac13\rho_r$, which was sufficient before and during the recombination era, but~is is almost negligible~now.

The $\Lambda$CDM model successfully describes numerous observations, including Type Ia
supernovae (SNe Ia) data, estimates of the Hubble parameter $H(z)$, manifestations of
baryonic acoustic oscillations (BAO), cosmic microwave background radiation (CMB), and
other data~\cite{Planck15,Planck18}. However this model does not explain the nature of
dark energy, the~small observable value of the phenomenological constant $\Lambda$ and
the approximate equality $\rho_x$ and $\rho_m$ now (although these densities evolve
differently).

Another essential problem in the $\Lambda$CDM model is the tension between Planck
estimations of the Hubble constant $H_0=67.37\pm0.54$ km\,/(s$\cdot$Mpc) \cite{Planck18}
(2018) and measurements of SH0ES  group in the Hubble Space Telescope
{%
 $H_0=73.2 \pm1.3$ km\,s${}^{-1}$ Mpc${}^{-1}$ \cite{Riess2020} (2020) or
$H_0=73.3 \pm1.04$ km\,s${}^{-1}$ Mpc${}^{-1}$ \cite{Riess2021} (2021).}

 The Planck estimations~\cite{Planck15,Planck18} are based on the $\Lambda$CDM model
and the Planck satellite measurements of the CMB anisotropy and power spectra related to
the early Universe at redshifts $z\sim1000$, whereas the SH0ES method  uses  local
distance ladder measurements of Cepheids in our Galaxy~\cite{Riess2020}  and in nearest
galaxies, in~particular, in~the Large Magellanic Cloud~\cite{HST19}, that implies $z$
close to 0 (the late Universe). This $H_0$ tension has not diminished during the last
years, and now it exceeds $4\sigma$.

Cosmologists suggested numerous scenarios for solving the mentioned problems with dark
energy and the $H_0$ tension; they include modifications of early or late dark energy,
dark energy with extra degrees of freedom, models with interaction in dark sector,
models with extra relativistic species, viscosity, modified gravity including $F(R)$
theories, and other models (see reviews
~\cite{BambaCNO:2012,NojiriO:2011,OdintsovO:2017,DiValentInter2020,DiValentRealmTens2021}
and papers
~\cite{KumarNunes2016,PanYDiVSCh2019,DiValentMMV2019,PanSharov2017,PanSharovYang2020,OdintsovSGSvisc2020,OdintsovSGS2017,OdintsovSGS2019,NojiriOSGS:2021,OdintsovSGS:2020}).

In this paper, we consider cosmological scenarios with modified equation of state (EoS)
for dark energy; they generalize the $\Lambda$CDM model and its simplest extensions: the
$w$CDM model with EoS
\begin{equation}
p_x=w\rho_x,\qquad w=\mbox{const}
   \label{wCDM}\end{equation}
 and the models with variable EoS
\begin{equation}
p_x=w(a)\,\rho_x,
   \label{war}\end{equation}
 where $w$ depends on the scale factor $a$. The~class (\ref{war})
includes the following well-known dark energy equations of state: the linear model
~\cite{Linear}
\begin{equation}
w = w_0+ w_1(a^{-1}-1) = w_0+ w_1z;
  \label{linear}\end{equation}
    {Chevallier}--Polarski--Linder (CPL) parametrization~\cite{ChP,Linder} 
\begin{equation}
w= w_0+ w_1(1-a) = w_0+ w_1 \frac{z}{1+z}
 \label{CPL}\end{equation}
 and (their generalization) Barboza--Alcaniz--Zhu--Silva (BAZS) EoS~\cite{BarbozaAZS09}
\begin{equation}
w= w_0+ w_1\frac{1-a^\beta}\beta = w_0+ w_1 \frac{1-(1+z)^{-\beta}}\beta.
 \label{BAZS}\end{equation}

 Here, the scale factor $a$ is normalized, so $a(t_0)=1$  at the present time $t_0$; $a$
is connected with redshift $z$:  $a=(1+z)^{-1}$. Obviously, BAZS  parametrization
(\ref{BAZS}) transforms into CPL EoS (\ref{CPL}) at $\beta=1$, into~the linear EoS
(\ref{linear}) if $\beta=-1$ and into the logarithmic EoS $w = w_0+ w_1\log(1+z),$
 if $\beta\to0$.

For all mentioned equations of state, one can integrate the continuity equation for
non-interacting dark energy:
\begin{equation}
\dot\rho_x = -3 H (\rho_x+p_x).
 \label{contDE}\end{equation}

In this paper, we explore the above scenarios and suggest the following generalization of
BAZS EoS:
\begin{equation}
w= w_0+ w_1\frac{1-a^\beta}\beta a^\gamma.
 \label{EoSn}\end{equation}

 We test these models, confronting them with the following observational data:  Type Ia
supernovae data (SNe Ia) from the Pantheon sample survey~\cite{Scolnic17}, data
extracted from Planck  2018 observations~\cite{Planck18,ChenHuangW2018} of cosmic
microwave background radiation (CMB) ,and estimations of the Hubble parameter $H(z)=\dot
a/a$ for different redshifts $z$ (the dot means $\frac {d}{dt}$). We use $H(z)$ data
from two sources: (a) from cosmic chronometers, that is, measured from differential ages
of galaxies, and (b) estimates of $H(z)$ obtained  from line-of-sight baryonic acoustic
oscillations (BAO) data.

This paper is organized as follows. In~the next section, we describe $H(z)$,  SNe Ia, and
CMB observational data analyzed here. Section~\ref{Models} is devoted to dynamics and free
model parameters for scenarios (\ref{wCDM})--(\ref{EoSn}). In~Section~\ref{Results} we
analyze the results of our calculations for these models, and 
estimate values of model
parameters including  the Hubble constant $H_0$, and in Section~\ref{Disc} we discuss the
results and their possible applications for alleviating the Hubble constant tension
problem.

\section{Observational~Data}
\label{Observ}

Observational data should be  described by the considered cosmological models. For~each
model, we calculate the best fit for its free parameters from the abovementioned data
sources: (a) Type Ia supernovae (SNe Ia) data from Pantheon sample~\cite{Scolnic17}, (b)
CMB data from Planck 2018~\cite{Planck18,ChenHuangW2018}, and (c) estimates of the
Hubble parameter $H(z)$ from cosmic chronometers and line-of-sight BAO~data.

The  Pantheon sample database~\cite{Scolnic17} for SNe Ia contains
$N_{\mbox{\scriptsize SN}}=1048$ data points of distance moduli $\mu_i^\mathrm{obs}$ at
redshifts $z_i$ in the range $0<z_i<1.92$. We compare them with theoretical values  by
minimizing the $\chi^2$ function:
\begin{equation}
\chi^2_{\mbox{\scriptsize SN}}(\theta_1,\dots)=\min\limits_{H_0}
\sum_{i,j=1}^{N_{\mbox{\scriptsize SN}}}
 \Delta\mu_i\big(C_{\mbox{\scriptsize SN}}^{-1}\big)_{ij} \Delta\mu_j,\qquad
 \Delta\mu_i=\mu^\mathrm{th}(z_i,\theta_1,\dots)-\mu^\mathrm{obs}_i\ ,
 \label{chiSN}\end{equation}

{Here,} $\theta_j$ are free model parameters, $C_{\mbox{\scriptsize SN}}$ is the covariance
matrix~\cite{Scolnic17}, and the~distance moduli {$\mu^\mathrm{th}$} are expressed via the %
luminosity distance $D_L$ depending on the spacial curvature fraction $\Omega_k$ and the
Hubble parameter $H(z)$:
  {%
 $$ \mu^\mathrm{th}(z) = m_B(z) -M_B = 5
\log_{10} \frac{D_L(z)}{10\mbox{pc}}, \qquad D_L (z)= (1+z)\, D_M,$$
 $$
D_M(z)= \frac{c}{H_0}S_k
 \bigg(H_0\int\limits_0^z\frac{d\tilde z}{H(\tilde
 z)}\bigg),\qquad S_k(x)=\left\{\begin{array}{ll} \sinh\big(x\sqrt{\Omega_k}\big)\big/\sqrt{\Omega_k}, &\Omega_k>0,\\
 x, & \Omega_k=0,\\ \sin\big(x\sqrt{|\Omega_k|}\big)\big/\sqrt{|\Omega_k|}. &
 \Omega_k<0,
 \end{array}\right.
 $$

Here, $m_B$ is a supernova apparent magnitude, and $M_B$ is its absolute magnitude. The~distance moduli {$\mu^\mathrm{th}$} are not Hubble-free and depend on the Hubble constant
$H_0= H(t_0)$ (via the summand $-5\log_{10}H_0$). On~the other hand,
$\mu_i^\mathrm{obs}$ is essentially connected with the  absolute magnitude $M_B$ and
calculated with corrections coming from deviations of lightcurve shape, SN Ia color,  and
mass of a host galaxy~\cite{Scolnic17,NesserisP2005}. In~the Pantheon sample, these
corrections and the connected pair $(H_0,M_B)$ were considered  as nuisance parameters,
in particular:
 ``Using only SNe, there is no constraint on $H_0$ since $H_0$ and $M_B$  are degenerate''~\cite{Scolnic17}.
We cannot divide uncertainties in the Hubble constant $H_0$ and possible  uncertainties
in $M_B$ \cite{Efstathiou2021,CamarenaM2021,NunesdV2021}.

Due to these reasons, we have to consider $H_0$ in Equation~(\ref{chiSN}) as a nuisance
parameter, its estimations cannot be obtained from $\chi^2_{\mathrm{SN}}$, and~we
minimize this function over  $H_0$~\mbox{\cite{OdintsovSGSvisc2020,OdintsovSGS2017,OdintsovSGS2019,NojiriOSGS:2021,OdintsovSGS:2020}.
 }}
However, SHe Ia data in
$\chi^2_{\mbox{\scriptsize SN}}$ is important for fitting other model~parameters.

{Unlike SNe Ia Pantheon  data, the~CMB observations are related to the photon-decoupling}
epoch near $z_*\simeq1090$. We use the following parameters extracted from Planck 2018
CMB {observations}~\cite{Planck18,ChenHuangW2018,NojiriOSGS:2021}:
\begin{equation}
\mathbf{x}=\left(R,\ell_A,\omega_b \right)\, ,\quad
R=\sqrt{\Omega_m^0}\frac{H_0D_M(z_*)}c\, ,\quad \ell_A=\frac{\pi D_M(z_*)}{r_s(z_*)}\, ,
\quad\omega_b=\Omega_b^0h^2
 \label{CMB} \end{equation}
   {%
 and their estimations for the non-flat $\Lambda$CDM + $\Omega_k$ model~\cite{ChenHuangW2018}:
\begin{equation}
\mathbf{x}^\mathrm{Pl}=\left( R^\mathrm{Pl},\ell_A^\mathrm{Pl},\omega_b^\mathrm{Pl}
\right) =\left( 1.7429\pm0.0051,\;301.409\pm0.091,\;0.0226\pm0.00017 \right).
\label{CMBpriors}
\end{equation}

{Considering} the flat case ($\Omega_k=0$) of these models, we use the flat $w$CDM data
~\cite{ChenHuangW2018}. }
The  comoving sound horizon $r_s$ at $z_*$ is calculated as the integral
\begin{equation}
 r_s(z_*)=  \int_{z_*}^{\infty}
\frac{c_s(\tilde z)}{H (\tilde z)}\,d\tilde
z=\frac1{\sqrt{3}}\int_0^{1/(1+z_*)}\frac{da}
 {a^2H(a)\sqrt{1+\big[3\Omega_b^0/(4\Omega_\gamma^0)\big]a}}
  \label{rs2}\end{equation}
 with  the fitting formula from Refs.~\cite{ChenHuangW2018,HuSugiyama95} for the value $z_*$.
The resulting $\chi^2$ function is
\begin{equation}
\chi^2_{\mathrm{CMB}}=\min_{\omega_b}\Delta\mathbf{x}\cdot C_{\mathrm{CMB}}^{-1}\left(
\Delta\mathbf{x} \right)^{T} ,\qquad \Delta
\mathbf{x}=\mathbf{x}-\mathbf{x}^\mathrm{Pl}\,, \label{chiCMB}
\end{equation}
    {%
 where we minimize over the normalized baryon fraction $\omega_b^0$ to diminish the effective
number $N_p$ of free model parameters.
  }
The covariance matrix $C_{\mathrm{CMB}}=\left\| \tilde C_{ij}\sigma_i\sigma_j \right\|$
and other details are described in papers~\cite{OdintsovSGSvisc2020,ChenHuangW2018}.

In this paper, we use the Hubble parameter data  $H(z)$ obtained from two different
sources
~\cite{OdintsovSGSvisc2020,OdintsovSGS2017,OdintsovSGS2019,NojiriOSGS:2021,OdintsovSGS:2020,SharovSin2020}.
The first one is the cosmic chronometers (CC), in~other words, estimations of $H(z)$ via
differences of ages $\Delta t$ for galaxies with close redshifts $\Delta z$ and
the formula
$$ 
 H (z)= \frac{\dot{a}}{a} 
  \simeq -\frac{1}{1+z} \frac{\Delta z}{\Delta t}.
 $$

{Here,} we include 31 CC $H(z)$ data points  from 
 Refs.~\cite{Simon,Stern,Moresco12,Zhang,Moresco15,Moresco16,Ratsimbazafy}  used earlier in
papers~\cite{OdintsovSGSvisc2020,OdintsovSGS2017,OdintsovSGS2019,NojiriOSGS:2021}
  and the recent estimate from Ref.~\cite{HzBorghi2021}; they are
shown in Table~\ref{HT}.

\begin{table}[th]  
{\caption{{$H(z)$ data from} cosmic chronometers (CC)
 \cite{Simon,Stern,Moresco12,Zhang,Moresco15,Moresco16,Ratsimbazafy,HzBorghi2021}
and line-of-sight   
BAO~\cite{Gaztanaga09,Blake12,ChuangW13,Chuang13,Busca13,Oka14,Font-Ribera14,Delubac15,Anderson14,Wang17,Alam17,Bautista17,Bourboux17,Zhu18,Zhao18}.
\label{HT}}
  \begin{tabular}{||c|c|c|l||c|c|c|l||}
   \hline
            \multicolumn{4}{|c||}{\textbf{CC Data}} & \multicolumn{4}{c||}{\boldmath\textbf{$H_\mathrm{BAO}$ Data}} \\ \hline
            \boldmath$z$ & \boldmath$H(z)$ &\boldmath$\sigma$ & \textbf{Refs} & \boldmath$z$ & \boldmath$H(z)$ & \boldmath$\sigma$ & \textbf{Refs} \\ \hline
            0.070 & 69 & 19.6  &Zhang\,14  & 0.240 & 79.69&2.992 & Gazta\~naga\,09\\ 
            0.090 & 69 & 12    &Simon\,05  & 0.30 & 81.7 & 6.22  & Oka\,14 \\ 
            0.120 & 68.6 &26.2 &Zhang\,14  & 0.31 & 78.18 & 4.74 & Wang\,17 \\ 
            0.170 & 83 & 8     &Simon\,05  & 0.34 & 83.8 & 3.66  & Gazta\~naga\,09\\ 
            0.1791 & 75 & 4    &Moresco\,12& 0.350 & 82.7 & 9.13 & ChuangW\,13\\ 
            0.1993 & 75 & 5    &Moresco\,12& 0.36  & 79.94 & 3.38 & Wang\,17 \\ 
            0.200 & 72.9 &29.6 &Zhang\,14  & 0.38 & 81.5 & 1.9   & Alam\,17\\ 
            0.270 & 77 & 14    &Simon\,05  & 0.400 & 82.04 &2.03 & Wang\,17 \\ 
            0.280 & 88.8 &36.6 &Zhang\,14  & 0.430 & 86.45&3.974 & Gazta\~naga\,09\\ 
            0.3519 & 83 & 14   &Moresco\,12& 0.44 & 82.6 & 7.8   & Blake\,12\\ 
            0.3802 & 83 & 13.5 &Moresco\,16& 0.44 & 84.81 & 1.83 & Wang\,17 \\ 
            0.400 & 95 & 17    &Simon\,05  & 0.48 & 87.79 & 2.03 & Wang\,17 \\ 
            0.4004 & 77 &10.2  &Moresco\,16& 0.51 & 90.4 & 1.9   & Alam\,17\\ 
            0.4247 & 87.1&11.2 &Moresco\,16& 0.52 & 94.35 & 2.64 & Wang\,17 \\ 
            0.445 & 92.8 &12.9 &Moresco\,16& 0.56 & 93.34 & 2.3  & Wang\,17 \\ 
            0.470 & 89 & 34    &Ratsimbazafy$\,$& 0.57 & 87.6 & 7.83  & Chuang\,13\\ 
            0.4783 & 80.9 & 9  &Moresco\,16& 0.57 & 96.8 & 3.4   & Anderson\,14\\ 
            0.48 & 97 & 62     &Stern\,10  & 0.59 & 98.48 &3.18  & Wang\,17 \\ 
            0.5929 & 104 & 13  &Moresco\,12& 0.600 & 87.9 & 6.1  & Blake\,12\\ 
            0.6797 & 92 & 8    &Moresco\,12& 0.61 & 97.3 & 2.1   & Alam\,17\\ 
            0.75   &98.8 &33.6 &Borghi\,21 & 0.64 & 98.82 & 2.98 & Wang\,17 \\ 
            0.7812 & 105 & 12  &Moresco\,12& 0.730 & 97.3 & 7.0  & Blake\,12\\ 
            0.8754 & 125 & 17  &Moresco\,12& 0.8 & 106.9 & 4.9   & Zhu\,18 \\ 
            0.880 & 90 & 40    &Stern\,10  & 0.978 &113.72&14.63 & Zhao\,19 \\ 
            0.900 & 117 & 23   &Simon\,05  & 1.0 & 120.7 & 7.3   & Zhu\,18 \\ 
            1.037 & 154 &20.17 &Moresco\,12& 1.230 &131.44&12.42 & Zhao\,19\\ 
            1.300 & 168 & 17   &Simon\,05  & 1.5 & 161.4 & 30.9  & Zhu\,18 \\ 
            1.363 & 160 & 33.6 &Moresco\,15& 1.526 &148.11&12.75 & Zhao\,19\\ 
            1.430 & 177 & 18   &Simon\,05  & 1.944& 172.63&14.79 & Zhao\,19\\ 
            1.530 & 140 & 14   &Simon\,05  & 2.0 & 189.9 & 32.9  & Zhu\,18 \\ 
            1.750 & 202 & 40   &Simon\,05  & 2.2 & 232.5 & 54.6; & Zhu\,18 \\ 
            1.965 & 186.5&50.4 &Moresco\,15& 2.300 & 224   & 8.57 & Buska\,13\\ 
              & & & &              2.330 & 224   & 8.0  & Bautista\,17\\ 
              & & & &              2.340 & 222   & 8.515& Delubac\,15\\ 
              & & & &              2.360 & 226   & 9.33 & Font-Ribera\,14$\!\!$\\ 
              & & & &              2.40  & 227.6 & 9.10 & Bourboux\,17\\
              \hline
        \end{tabular}}
\end{table}
\unskip

These 32 CC data points need a covariance matrix of systematic
uncertainties connected with a choice of initial mass function, metallicity, star
formation history, stellar population synthesis models, and other factors
~\cite{Moresco2020,Moresco2021}.

 We describe these uncertainties as corrections $ \Delta C_{\mathrm{H}}$ to the diagonal
covariance matrix $C^{\mathrm{d}}_{\mathrm{H}}=\mbox{diag}\{\sigma_i^{-2}\}$ (from
Table~\ref{HT}) taking into account their diagonal terms in the form~\cite{Moresco2020}
$$
(\Delta C_{\mathrm{H}})_{ii}=\big[\eta(z_i)\,H(z_i)\big]^2.
$$

{Here,} $\eta(z)$ is  a mean percentage bias depending on redshift $z$. We consider ``the
best-case scenario'' from the paper~\cite{Moresco2020} for $\eta(z)$ and include these
contributions of stellar population synthesis and  metallicity omitting the non-diagonal
terms $\Delta (C_{\mathrm{H}})_{ij}$ (they are negligible for metallicity
~\cite{Moresco2021}).

The second source of $H(z)$ estimates is the baryon acoustic oscillation (BAO) data
along the line-of-sight direction. We use here 36 $H_{\mathrm{BAO}}(z)$ data points from
Refs.~\cite{Gaztanaga09,Blake12,ChuangW13,Chuang13,Busca13,Oka14,Font-Ribera14,Delubac15,Anderson14,Wang17,Alam17,Bautista17,Bourboux17,Zhu18,Zhao18}
 (see  Table~\ref{HT}). They were considered earlier in
Ref.~\cite{OdintsovSGS:2020}.

 Some of the $H(z)$
measurements~\cite{Gaztanaga09,Blake12,ChuangW13,Chuang13,Busca13,Oka14,Font-Ribera14,Delubac15,Anderson14,Wang17,Alam17,Bautista17,Bourboux17,Zhu18,Zhao18}
  in Table~\ref{HT} used the same or overlapping
large-scale structure data, so these $H$ estimates for close redshifts $z$ may be in
duplicate.
 It concerns, for~example, measurements of  Delubac~et~al. \cite{Delubac15},
Font-Ribera et al. \cite{Font-Ribera14}, uses quasars with Lyman-$\alpha$ forest from
Data Release 11 SDSS-III survey; estimates of Alam~et~al. \cite{Alam17}, Wang~et~al.
\cite{Wang17}, Bautista~et~al.~\cite{Bautista17}, and Bourboux~et~al.~\cite{Bourboux17}
were made with data from or DR12 of SDSS-III etc. To~avoid this doubling, we multiply
the errors $\sigma_i$ by  $\sqrt2$ for $H_{\mathrm{BAO}}$ estimates in Table~\ref{HT}
with close $z$, data, and~methods.

Note that $H_{\mathrm{BAO}}$ estimates in Table~\ref{HT} should be multiplied by the
factor $r^{\mathrm{fid}}_d/r_d$, where fiducial values $r^{\mathrm{fid}}_d$ of the sound
horizon size $r_d=r_s(z_d)$ at the drag epoch vary from 147.33 Mpc to  157.2 Mpc for
different
authors~\cite{Gaztanaga09,Blake12,ChuangW13,Chuang13,Busca13,Oka14,Font-Ribera14,Delubac15,Anderson14,Wang17,Alam17,Bautista17,Bourboux17,Zhu18,Zhao18}.
%
%
%
 We include this correction to errors  $\sigma_i$ quadratically, comparing deviations of
$r^{\mathrm{fid}}_d$  with $r_s(z_d)$ calculated with Formula (\ref{rs2}) for a
considered~model.

  For any cosmological model we calculate the $\chi^2$ function
\begin{equation}
    \chi_H^2(\theta_1,\dots)=\sum_{j=1}^{N_H}\bigg[\frac{H(z_j,\theta_1,\dots)-H^{obs}(z_j)}{\sigma _j}  \bigg]^2
    \label{chiH}
 \end{equation}
  by using (a) only CC $H(z)$ data 
   and (b) the full set CC
+ $H_\mathrm{BAO}$ data. Note that $H_\mathrm{BAO}$ data points are correlated with BAO
angular distances considered in the previous papers~\mbox{\cite{OdintsovSGS2017,OdintsovSGS2019,NojiriOSGS:2021}}). Thus, here, we do not use data with
BAO angular distances, to avoid any~correlation.

\section{Models}
\label{Models}

We explore all considered models in a homogeneous isotropic universe with the
Friedmann--Lema\^itre--Robertson--Walker metric
 $$ 
ds^2 = -dt^2 + a^2 (t) \left[\frac{dr^2}{1-k r^2} + r^2 \left(d \theta^2 + \sin^2 \theta\,
d \phi^2\right)  \right],
 $$ 
 where $k$ is the sign of spatial curvature.
 In this case, the Einstein equations are reduced to the system of the Friedmann equation
\begin{equation}
    3\frac{\dot{a}^2+k}{a^2}=8\pi G\rho
\label{Fried}
\end{equation}
and the continuity equation
\begin{equation}
    \dot\rho+3H(\rho+p)=0.
    \label{cont}
\end{equation}

{Here,} the total density $\rho$ includes densities of the abovementioned cold
pressureless matter (dark matter unified with baryonic matter), radiation, and dark
energy:
\begin{equation}
    \rho=\rho_m+\rho_r+\rho_x,\qquad \rho_m=\rho_c+\rho_b.
    \label{rho}\end{equation}

 We suppose here that dark energy and the mentioned components do
not interact in the form~\cite{PanYDiVSCh2019,DiValentMMV2019,PanSharov2017,PanSharovYang2020} and independently
satisfy the continuity Equation~(\ref{contDE}) or  (\ref{cont}). We integrate this
equation for cold and relativistic matter:
\begin{equation}
    \rho_m=\rho_m^0 a^{-3},\qquad
     \rho_r=\rho_r^0a^{-4}
    \label{rhomr}
\end{equation}
(the index ``0'' corresponds to the present time $t_0$) and  substitute these relations
into the Friedmann equation~(\ref{Fried}) that can be rewritten as
\begin{equation}
H^2 = H_0^2 \Big(\Omega_m^0 a^{-3}+\Omega_r^0a^{-4}+\Omega_ka^{-2} +\Omega_x(a)\Big)
 \label{Ha}\end{equation}
or
\begin{equation}
H(z) = H_0 \sqrt{\Omega^0_m(1+z)^3+\Omega_r^0(1+z)^4+\Omega_k(1+z)^2 +\Omega_x(z)}.
 \label{Hz}
\end{equation}
 {Here,}
 $$ \Omega^0_j = \frac{8\pi G\rho^0_j}{3H^2_0}, \quad j=m,r,\qquad
  \Omega_k = \frac{-k}{a^2_0H^2_0},\qquad
 \Omega_x(a) = \frac{8\pi G\rho_x(a)}{3H^2_0}. $$

The dark energy fraction $\Omega_x(a)$ results from  the continuity
Equation~(\ref{contDE}) $\dot\rho_x+3H(\rho_x+p_x)=0$, that, for the variable EoS (\ref{war}) $
_x=w(a)\,\rho_x$, is reduced to the form
\begin{equation}
\log\Omega_x(a) =-3 \int\big[1+w(a)\big]\,a^{-1}da.
 \label{cont2}
\end{equation}

In particular, for~BAZS  parametrization (\ref{BAZS}) $w= w_0+ w_1(1-a^\beta)/\beta $
\cite{BarbozaAZS09}, the expression (\ref{cont2}) is
\begin{equation}
 \Omega_x(a) =  \Omega_x^0\,a^{-3(1+w_0+ w_1/\beta)}\exp\bigg[\frac{3w_1(a^\beta-1)}{\beta^2}\bigg].
 \label{OmBAZS}\end{equation}
 {One} should substitute it into Equation~(\ref{Ha}).

In the above--mentioned particular cases, the BAZS formula (\ref{OmBAZS}) takes the form
\begin{equation}
 \Omega_x(a) =  \Omega_x^0\,a^{-3(1+w_0- w_1)} e^{3w_1z},
 \label{Omlin}\end{equation}
 for the linear model (\ref{linear}) $w= w_0+ w_1z$ (if $\beta=-1$) and
\begin{equation}
 \Omega_x(a) =  \Omega_x^0\,a^{-3(1+w_0+ w_1)}e^{3w_1(a-1)}.
 \label{OmCPL}\end{equation}
 for  CPL EoS (\ref{CPL}) $w= w_0+ w_1(1-a)$ (if $\beta=1$). In~the case $w_1=0$
(and $w_0\equiv w$), both models (\ref{Omlin}) and (\ref{OmCPL}) transform into the
$w$CDM model with
\begin{equation}
 \Omega_x(a) =  \Omega_x^0\,a^{-3(1+w)},
 \label{OmwCDM}\end{equation}
 Its particular case at $w=-1$ is the $\Lambda$CDM  model where $\Omega_x = \Omega_x^0=
 \Omega_\Lambda={}$const.

For the generalization (\ref{EoSn}) (with the factor $\gamma$) of  BAZS  parametrization
(\ref{BAZS}), the expression (\ref{cont2}) takes the form
\begin{equation}
 \Omega_x(a) =  \Omega_x^0\,a^{-3(1+w_0)}
 \exp\bigg[3\frac{w_1}\beta\bigg(\frac{1-a^\gamma}\gamma+\frac{a^{\beta+\gamma}-1}{\beta+\gamma}\bigg)\bigg].
 \label{OmEoSn}\end{equation}
{If} $\gamma\to0$ it transforms into Equation~(\ref{OmBAZS}), and in the case  $\beta\to0$ we
have
\begin{equation}
 \Omega_x(a) =  \Omega_x^0\,a^{-3(1+w_0-w_1a^\gamma/\gamma)}
 \exp\bigg(3w_1\frac{1-a^\gamma}{\gamma^2}\bigg).
 \label{Ombe0}\end{equation}

For all considered models, the dark energy fraction $\Omega_x^0$ and other  $\Omega_j$
satisfy the equality
$$\Omega^0_m + \Omega^0_r+\Omega_k  + \Omega^0_x = 1, $$
 resulting from  Equation~(\ref{Ha}) or (\ref{Hz}). Further, if~we fix the ratio
~\cite{OdintsovSGS2017,OdintsovSGS2019,NojiriOSGS:2021,Planck13}
\begin{equation}
 X_r = \frac{\rho^0_r}{\rho^0_m}=\frac{\Omega^0_r}{\Omega^0_m}=2.9656 \cdot 10^{-4}
\label{Xr}\end{equation}
  {%
  to diminish the number $N_p$ of free model parameters.
 }
  We will work with the following five free parameters in the linear (\ref{Omlin}) and CPL (\ref{OmCPL})
 models:
\begin{equation}
 \Omega^0_m,\quad \Omega_k,\quad H_0,\quad w_0,\quad w_1 .
\label{5param}\end{equation}
 {Here,} we consider $\omega_b$ in $\chi^2_{\mathrm{CMB}}$ (\ref{chiCMB}) as a nuisance parameter.
 In the $\Lambda$CDM  model the number of parameters $N_p=3$ ($\Omega^0_m$, $\Omega_k$, $H_0$), in the $w$CDM model (\ref{OmwCDM}) $N_p=4$ with $w_0\equiv w$, in the  generalized model (\ref{EoSn}), (\ref{OmEoSn}) we have $N_p=7$  free parameters
 with additional $\beta$ and $\gamma$ to the set (\ref{5param}).

 In the next sections, we compare predictions of these models with the
observational data from Section~\ref{Observ}.

\section{Results}
\label{Results}

We evaluate how the considered models fit the observations taking into account the
$\chi^2$ functions for SNe Ia (\ref{chiSN}), CMB (\ref{chiCMB}), and  $H(z)$ data
(\ref{chiH}) in the form
\begin{equation}
  \chi^2_\mathrm{tot}=\chi^2_\mathrm{SN}+\chi^2_H+\chi^2_{\mathrm{CMB}}.
 \label{chitot} \end{equation}

{For} the Hubble parameter data,  we separately use (a) only 32  data points with cosmic chronometers (CC)
$H(z)$ data and (b) the full set CC + $H_\mathrm{BAO}$ data (see  Table~\ref{HT}).

\vspace{-3pt}

 \begin{figure}[bh]
 \hspace{-0.5cm}
\includegraphics[width=13.0 cm]{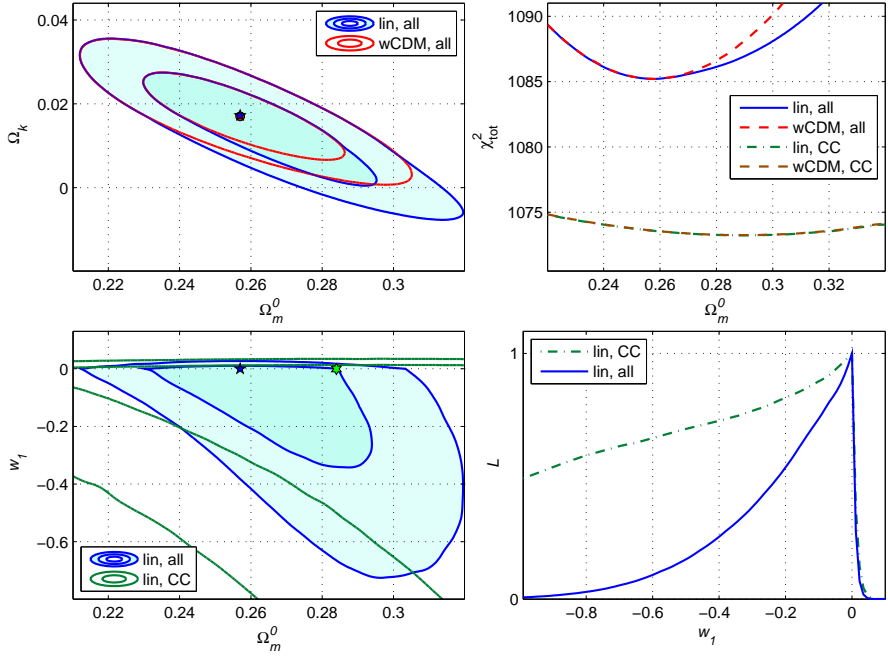}
\caption{{For the linear model} (\ref{linear}), $1\sigma$, $2\sigma$ contour plots in
the left panels (``all'' means all $H(z)$ data); one-parameter distributions
$\chi^2_\mathrm{tot}(\Omega_m^0)$ and likelihood functions ${\cal L}(w_1)$ in the right
panels are compared with the $w$CDM model. The stars and circles denote positions of
$\chi^2_\mathrm{tot}$ minima points.}
 \label{F1lin}
\end{figure}

When we compare two models with different  number $N_p$ of free parameters, we can
expect that the model with larger $N_p$ achieves more success in minimizing
$\chi^2_\mathrm{tot}$. However,  some models are not successful in this sense. In~particular, the~linear model (\ref{linear}), (\ref{Omlin}) with $N_p=5$ parameters
(\ref{5param}), with our set of observational data, yields the same minimal value
$\min\chi^2_\mathrm{tot}\simeq1092.34$ (for  CC + $H_\mathrm{BAO}$) as the $w$CDM model
(\ref{OmwCDM}) with $N_p=4$. The~reason is the following: the best fitted value $w_1$
for the linear model (\ref{linear}) is very close to zero ($w_1\simeq-0.0006$). In~this
case, the linear model works as the $w$CDM model (\ref{OmwCDM}) and yields the same
 $\min\chi^2_\mathrm{tot}$.

Such a behavior of the model (\ref{linear}) is shown in Figure~\ref{F1lin}, where
$1\sigma$ and $2\sigma$ filled contour plots in the left panels correspond to the full
set of  $H(z)$ data (here and below, ``all'' denotes CC + $H_\mathrm{BAO}$). The~$1\sigma$, $2\sigma$ contours for the $w$CDM model (red lines) in the
$\Omega_m^0-\Omega_k$ plane behave similarly and closely; positions of
$\chi^2_\mathrm{tot}$ minima point (shown as the star and the circle) practically
coincide.

Here, the contours are drawn for $\chi^2$ functions minimized over all other parameters,
in particular, for~the linear model (\ref{linear}) in the top-left panel:
\begin{equation}
 \chi^2_\mathrm{tot}(\Omega_m^0, \Omega_k)=\min\limits_{H_0,w_0,w_1} \chi^2_\mathrm{tot}.
 \label{chi2p}
\end{equation}

 A similar picture also takes place for only CC $H(z)$
data, where $\min\chi^2_\mathrm{tot}\simeq1074.73$ for both models. Equality of these
minima for only CC and all  $H(z)$ data is illustrated with one-parameter distributions
$\chi^2_\mathrm{tot}(\Omega_m^0)$ in the top-right panel. In~one-parameter distributions,
we also minimize over all other~parameters.

In the $\Omega_m^0-w_1$ plane (the bottom-left panel of Fig.~\ref{F1lin}) we see that for both models, minima
of $\chi^2_\mathrm{tot}$ are achieved near $w_1=0$. It is also shown in the bottom-right
panel, where the likelihood functions
\begin{equation}
{\cal L}(w_1)\sim\exp\big[-\chi^2_\mathrm{tot}(w_1)/2\big]
  \label{likeli}
\end{equation}
 are depicted for the linear~model.

\begin{figure}[bh]
\centering
\includegraphics[width=16 cm]{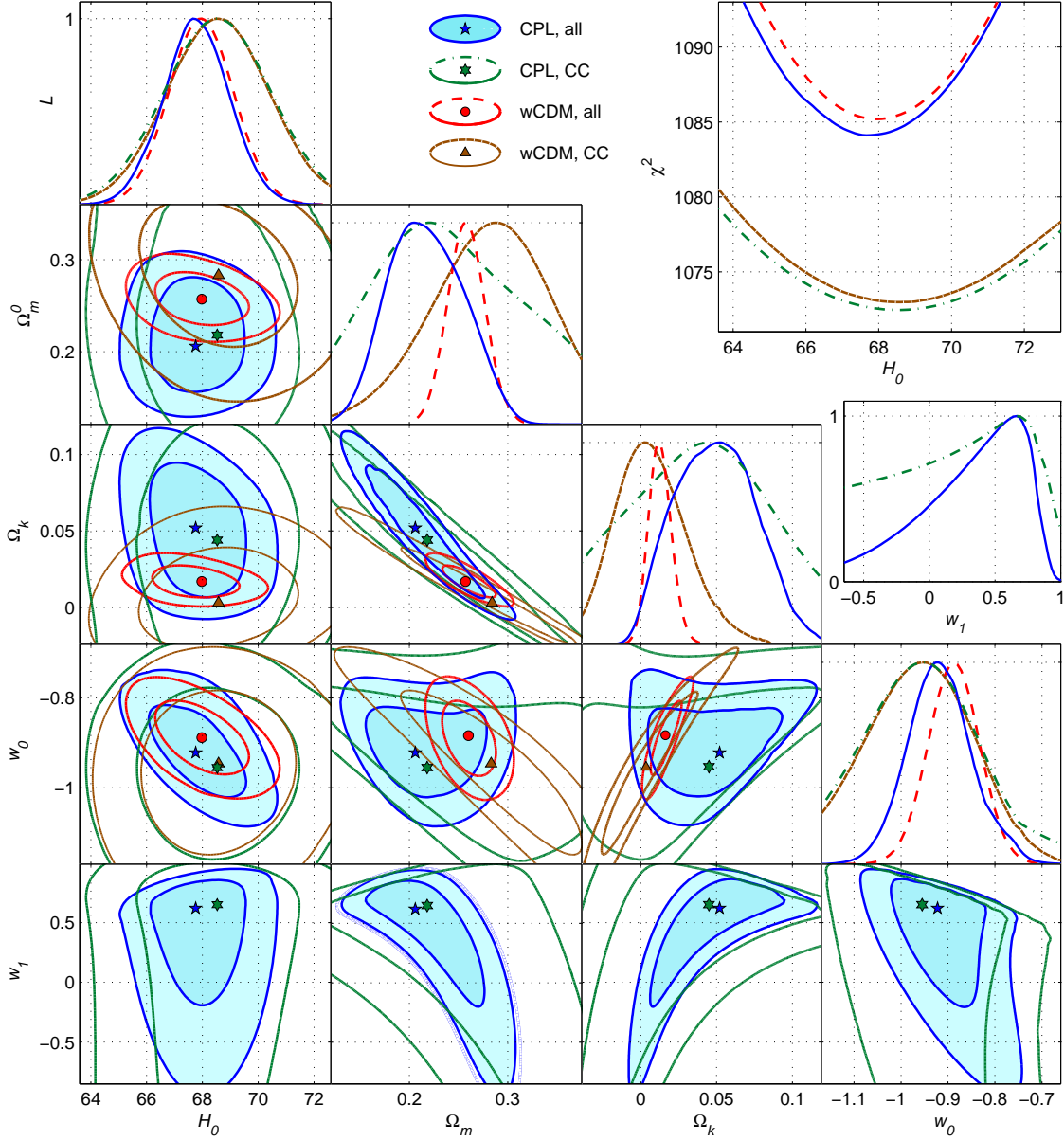}
\caption{CPL model (\ref{OmCPL}) in comparison with the $w$CDM model:  $1\sigma$,
$2\sigma$ contours, likelihoods ${\cal L}(\theta_j)$, and one-parameter distributions
$\chi^2_\mathrm{tot}(H_0)$.\label{F2CPL}}
\end{figure}

For the same observational data we can observe in  Figure~\ref{F2CPL} more successful
behavior of the Chevallier--Polarski--Linder (CPL) model (\ref{OmCPL})
\cite{ChP,Linder}.
 Here, $1\sigma$ and $2\sigma$ contours are drawn for
$\chi^2_\mathrm{tot}(\theta_i, \theta_j)$ of the type (\ref{chi2p}) in all planes of two
parameters $(\theta_i-\theta_j)$ in notation of  Figure~\ref{F1lin}. We consider four
cases: for two models (CPL and $w$CDM) we calculate $\chi^2_\mathrm{tot}$ for CC and all
$H(z)$ data, positions of all $\chi^2_\mathrm{tot}$ minima points are shown. Naturally,
in~the panels with $w_1$, only the CPL model is~presented.

The likelihood functions ${\cal L}(\theta_i)$ of the type (\ref{likeli}) are shown in
Figure~\ref{F2CPL} for all five model parameters (\ref{5param}). They are used for
estimating the best fits and $1\sigma$ errors for these parameters, summarized below in
Table~\ref{Estim}.

The CPL model achieves lower values of $\min\chi^2_\mathrm{tot}$ in comparison with the
$w$CDM and $\Lambda$CDM models. It can be seen in Table~\ref{Estim} and in the top-right
panel of Figure~\ref{F2CPL}, where one-parameter distributions $\chi^2_\mathrm{tot}(H_0)$
of these models are compared. The~graphs $\chi^2_\mathrm{tot}(H_0)$ and the
correspondent likelihoods ${\cal L}(H_0)$ in the top-left panel demonstrate that the
best fitted values  $H_0$ are very close for the $w$CDM and CPL models and differ more
essentially when we compare CC and all $H(z)$ data.

The best fits of $\Omega_m^0$ depend stronger on the chosen model and vary from
$\Omega_m^0\simeq0.206$ for CPL, all $H(z)$ to $\Omega_m^0\simeq0.289$ for $w$CDM, CC.
The best fits of $\Omega_k$ behave similarly, but~with the maximal estimate for CPL, all
$H(z)$. The~best CPL fits for $w_1$ are close to $0.66$ for both variants of $H(z)$ data.
This value is far from zero; in~other words, the CPL model with the considered
observational data behaves differently to the $w$CDM~model.

 \begin{figure}[ht]
\includegraphics[width=13.0 cm]{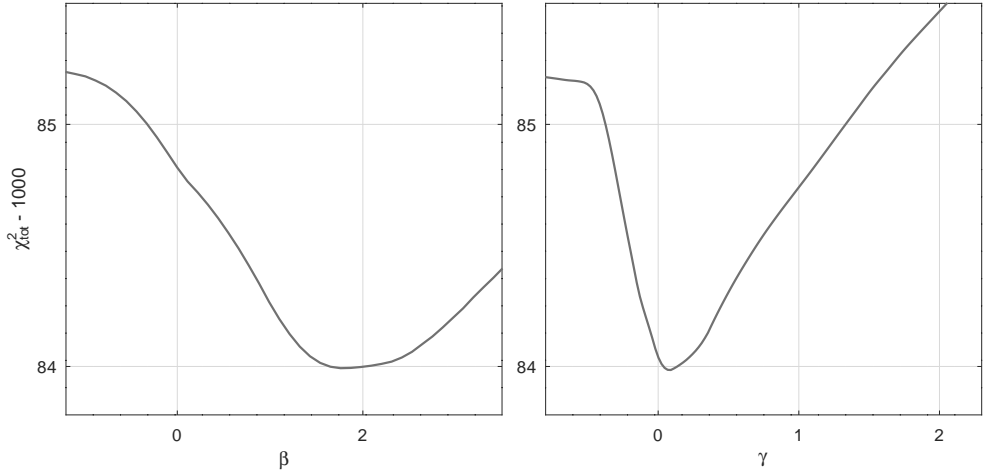}
\caption{The generalized  model (\ref{OmEoSn}) with all $H(z)$ data:
one-parameter distributions for $\beta$ and $\gamma$.}
 \label{F3gen}
\end{figure}

The CPL model achieves good results, but~it is not the most successful scenario in the
frameworks of the generalized model (\ref{EoSn}), (\ref{OmEoSn}). However, this
generalized model has $N_p=7$ free parameters, including $\beta$ and $\gamma$ in
addition to five parameters (\ref{5param}). This large number is the serious disadvantage
of the generalized model  (\ref{OmEoSn}) if we keep in mind informational criteria
~\cite{OdintsovSGS2019,NojiriOSGS:2021,OdintsovSGS:2020} and difficulties in
calculations.


  {%
Following these reasons, we calculated the  $\chi^2_\mathrm{tot}$ function (\ref{chitot})
for the generalized model  (\ref{Ha}), (\ref{OmEoSn}) with all $H(z)$ data searching its
minimum in the $\beta-\gamma$ plane. The~results are presented in Figure~\ref{F3gen} as
 one-parameter distributions
$\chi^2_\mathrm{tot}(\beta)$ and $\chi^2_\mathrm{tot}(\gamma)$ (minimized over all other parameters).
 }

\begin{table}[hb]
\caption{The best fit values with $1\sigma$ estimates of model parameters and
$\min\chi^2_\mathrm{tot}$ for SNe Ia + CMB and CC or all $H(z)$ data for the models:
$\beta=2$ (\ref{be2}), CPL (\ref{OmCPL}),  $w$CDM, and  $\Lambda$CDM. \label{Estim}}
\begin{tabular}{||c|c|c|c|c|c|c|c||}
\hline
\textbf{Model}  & \textbf{Data} & \boldmath$\min\chi^2_\mathrm{tot}/d.o.f$
& \boldmath$H_0$ & \boldmath$\Omega_m^0$  & \boldmath$\Omega_k$ &
\boldmath$w_0\equiv w$ & \boldmath$w_1$ \\ \hline
 $\beta=2$ & CC & $1072.30/1079$  & $68.40^{+1.86}_{-1.91}$ & {$0.177^{+0.113}_{-0.072}$} & $0.080^{+0.067}_{-0.076}$ & {$-0.967^{+0.12}_{-0.14}$} & {$1.33^{+0.42}_{-1.56}$}\rule{0pt}{1.4em} \\
\hline
  CPL   & CC & $\!\!1072.45/1079\!\!$ &$\!\!68.53^{+1.84}_{-1.88}\!\!$ & {$\!\!0.218^{+0.102}_{-0.074}\!\!$} & $\!\!0.044^{+0.047}_{-0.059}\!\!$ & {$\!-0.955^{+0.14}_{-0.14}\!$} & {$\!\!0.65^{+0.25}_{-1.04}\!\!$}\rule{0pt}{1.4em} \\
\hline
 $\!\!w$CDM$\!\!$& CC &  $\!\!1072.97/1080\!\!$ & $\!\!68.58^{+1.85}_{-1.89}\!\!$ & {$\!\!0.283^{+0.055}_{-0.058}\!\!$} & $\!\!0.003^{+0.024}_{-0.021}\!\!$ & {$\!\!-0.946^{+0.14}_{-0.14}\!\!$} & - \rule{0pt}{1.4em} \\
\hline
$\!\!\Lambda$CDM$\!\!$&  CC & $\!\!1073.81/1081\!\!$ & $\!\!68.52^{+1.89}_{-1.94}\!\!$ &{$\!\!0.305^{+0.029}_{-0.029}\!\!$} &  $\!\!0.0052^{+0.0018}_{-0.0018}\!\!$ & $-1$&  - \rule{0pt}{1.4em}  \\
\hline
 $\beta=2$ & all & $\!\!1084.03/1115\!\!$ & $\!\!67.65^{+1.28}_{-1.26}\!\!$ & {$\!\!0.190^{+0.071}_{-0.072}\!\!$} & $\!\!0.068^{+0.072}_{-0.053}\!\!$ & {$\!\!-0.96^{+0.087}_{-0.067}\!\!$}&  {$\!\!1.12^{+0.40}_{-1.09}\!\!$}\rule{0pt}{1.4em} \\
\hline
  CPL   & all & $\!\!1084.28/1115\!\!$ & $\!\!67.76^{+1.15}_{-1.01}\!\!$ &  {$\!\!0.206^{+0.054}_{-0.025}\!\!$} & $\!\!0.052^{+0.027}_{-0.036}\!\!$ & {$\!\!-0.922^{+0.07}_{-0.06}\!\!$}&  {$\!\!0.62^{+0.18}_{-0.52}\!\!$}\rule{0pt}{1.4em} \\
\hline
 $\!\!w$CDM$\!\!$& all & $\!\!1085.19/1116\!\!$ & $\!\!67.98^{+1.11}_{-1.11}\!\!$ &  {$\!\!0.257^{+0.019}_{-0.019}\!\!$} & $\!\!0.017^{+0.008}_{-0.008}\!\!$ & {$\!\!-0.89^{+0.05}_{-0.05}\!\!$}& - \rule{0pt}{1.4em} \\
\hline
$\!\!\Lambda$CDM$\!\!$& all &  $\!\!1089.42/1117\!\!$ & $\!\!69.02^{+1.04}_{-1.06}\!\!$ & {$\!\!0.272^{+0.013}_{-0.012}\!\!$} &  $\!\!0.004^{+0.0017}_{-0.0017}\!\!$ & $-1$&  - \rule{0pt}{1.4em}  \\
\hline
\end{tabular}
\end{table}

\begin{figure}[bh]
\centering
\includegraphics[width=16 cm]{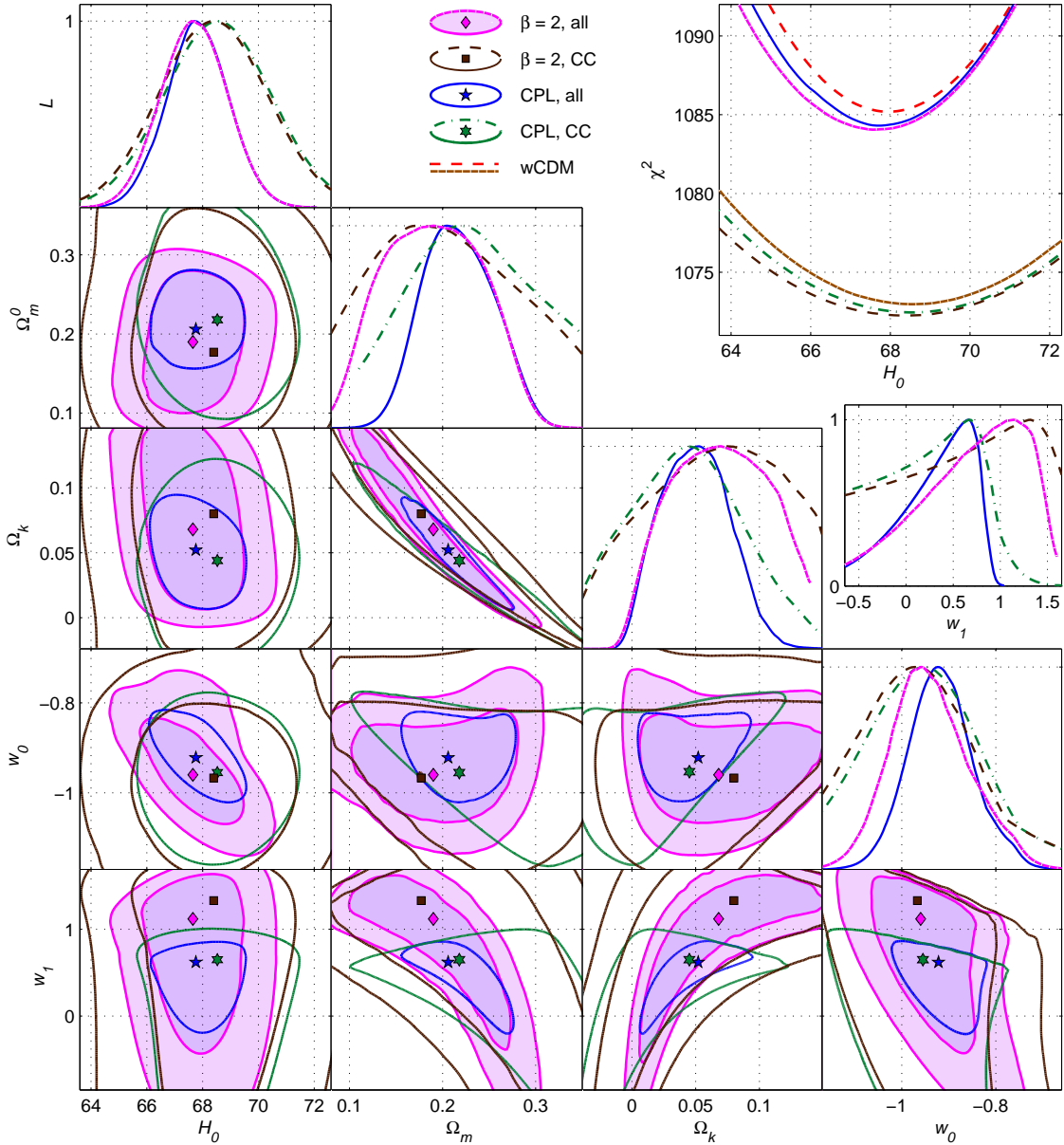}
 \caption{BAZS model $\beta=2$ (\ref{be2}) with $1\sigma$, $2\sigma$ contours in
comparison with CPL model ($1\sigma$ contours), likelihoods ${\cal L}(\theta_j)$, and
one-parameter distributions $\chi^2_\mathrm{tot}(H_0)$.
 \label{F4be2}}
\end{figure}

We see in Figure~\ref{F3gen} that the absolute minimum for the generalized  model
(\ref{OmEoSn}) $\min\chi^2_\mathrm{tot}\simeq 1084.0$ is achieved near the point
$\beta=2$, $\gamma=0$. One may conclude that the Barboza--Alcaniz--Zhu--Silva (BAZS) model
~\cite{BarbozaAZS09} (corresponding to  $\gamma=0$) with $\Omega_x(a)$ (\ref{OmBAZS}) and
$\beta=2$ appeared to be the most successful for the considered observational data Ia +
CMB + CC + $H_{\mathrm{BAO}}$. When we fix $\beta=2$, the~BAZS model with
\begin{equation}
w= w_0+ w_1(1-a^2)/2
 \label{be2}\end{equation}
 will have five free model parameters (\ref{5param}). We investigate this model (denoted below as ``$\beta=2$'') in detail;
 the results are presented in Table~\ref{Estim} and in Figure~\ref{F4be2}.

Figure~\ref{F4be2} illustrates the BAZS model with $\beta=2$ (\ref{be2}): $1\sigma$ and
$2\sigma$ contour plots are shown for all $H(z)$ data (filled contours) and for only CC
data. They are compared with the corresponding $1\sigma$ contours of the CPL model
(shown also in Figure~\ref{F2CPL}). One can compare the related one-parameter
distributions $\chi^2_\mathrm{tot}(H_0)$ in the top-right panel and likelihood functions
(\ref{likeli}) ${\cal L}(H_0)$, ${\cal L}(\Omega_m^0)$, etc.

Figure~\ref{F4be2} and Table~\ref{Estim} demonstrate that the  $\beta=2$ BAZS model
(\ref{be2}) is more successful in minimizing $\chi^2_\mathrm{tot}$ if we compare it with the
CPL model (and, naturally, with~the $w$CDM and  $\Lambda$CDM models). For~only CC data
and for all $H(z)$ data, the best fit values of $\chi^2_\mathrm{tot}$ for the $\beta=2$
model (\ref{be2}) are achieved at lower values of $\Omega_m^0$
($\Omega_m^0=0.177^{+0.113}_{-0.072}$ for CC and $0.19^{+0.071}_{-0.072}$ for all $H$
data) and  at larger values of $\Omega_k$ in comparison with CPL model. However,  the~best
fit values of $H_0$  approximately coincide for all considered models; they depend on a
chosen dataset: only CC data or all $H(z)$ data.

The best fitted values of $w_0$ are close for the $\beta=2$, CPL, and  $w$CDM models in
the CC case and slightly differ for all $H(z)$ data. The~optimal values of $w_1$ are
larger for the $\beta=2$  model  ($w_1=1.33^{+0.42}_{-1.56}$ for CC and
$1.12^{+0.40}_{-1.09}$ for all $H$ data) if we compare with the CPL model; however, one
should take into account the factor $\frac12$ in EoS (\ref{be2}): $w= w_0+
w_1(1-a^2)/2$.

To compare models with different number $N_p$ of free model parameters, we use here the
Akaike information criterion~\cite{OdintsovSGS2019,NojiriOSGS:2021,Akaike74}:
 \begin {equation}
\mathrm{AIC} = \min\chi^2_{tot} +2N_p.
  \label{Akaike}
\end{equation}

 {This} criterion emphasizes the advantage of models with small number of $N_p$. It can be seen in Table~\ref{EstAIC}
for the mentioned models, where, for only CC $H(z)$ data, the minimal  Akaike values
(\ref{Akaike}) are achieved for the $\Lambda$CDM ($\mathrm{AIC} = 1079.81$) and the
$w$CDM ($\mathrm{AIC} = 1080.97$) models.
  {%
 However, for~all $H(z)$ data, the $\beta=2$
model (\ref{be2}) with $N_p=5$ parameters appeared to be more successful than $\Lambda$CDM  not only in
$\min\chi^2_\mathrm{tot}\simeq 1084.03$, but~also with Akaike information (\ref{Akaike}):
$\mathrm{AIC} = 1094.03$. The~ lowest $\mathrm{AIC} = 1093.19$ is achieved here for  the
$w$CDM~model.
 }

On the right side of the  Table~\ref{EstAIC}, we present the predictions of the same
models on their flat case $k=0$ ($\Omega_k=0$). In~this case, the values
$\min\chi^2_\mathrm{tot}$ and AIC appeared to be larger, and~ the $\beta=2$ model
(\ref{be2}) loses its advantage in comparison with CPL scenario for all  $H(z)$~data.

\begin{table}[ht]
\small \caption{The best fits of  $\min\chi^2_\mathrm{tot}$,  AIC, and  $H_0$  for the
models $\beta=2$ (\ref{be2}), CPL (\ref{OmCPL}), $w$CDM, $\Lambda$CDM, and these models
in the spatially flat case ($k=0$). \label{EstAIC}}
\begin{tabular}{||c|c|c|c|c|c|c|c|c||} \hline
\textbf{Model}  & \textbf{Data} & \boldmath$\min\chi^2_\mathrm{tot}$ & \textbf{AIC}
& \boldmath$H_0$ & \textbf{Flat Model} & \boldmath$\min\chi^2_\mathrm{tot}/d.o.f$ & \textbf{AIC} & \boldmath$H_0$ \\
\hline
 $\beta=2$ & CC & $1072.30$ &$1082.30$ & $68.40^{+1.86}_{-1.91}$ & {flat $\beta=2$} & $!1072.99/1080$
 &$1080.99$  & $68.46^{+1.82}_{-1.81}$\rule{0pt}{1.4em} \\
\hline
  CPL  & CC & $\!\!1072.45\!\!$ &$\!\!1082.45\!\!$&$\!\!68.53^{+1.84}_{-1.88}\!\!$ & {$\!\!$flat CPL$\!\!$} & $\!\!1072.98/1080\!\!$ &$\!\!1080.98\!\!$ & $\!\!68.47^{+1.81}_{-1.80}\!\!$\rule{0pt}{1.4em} \\
\hline
 $\!\!w$CDM$\!\!$ & CC &  $\!\!1072.97\!\!$ &$\!\!1080.97\!\!$& $\!\!68.58^{+1.85}_{-1.89}\!\!$ & {$\!\!$flat $w$CDM$\!\!$} & $\!\!1073.01/1081\!\!$ &$\!\!1079.01\!\!$ &$\!\!68.50^{+1.81}_{-1.79}\!\!$\rule{0pt}{1.4em} \\
\hline
$\!\!\Lambda$CDM$\!\!$&  CC & $\!\!1073.81\!\!$ &$\!\!1079.81\!\!$& $\!\!68.52^{+1.89}_{-1.94}\!\!$ &{$\!\!$flat $\Lambda$CDM$\!\!$} & $\!\!1073.93/1082\!\!$ &$\!\!1077.93\!\!$& $\!\!69.21^{+1.92}_{-1.91}\!\!$ \rule{0pt}{1.4em}  \\
\hline
 $\beta=2$ & all & $\!\!1084.03\!\!$ &$\!\!1094.03\!\!$& $\!\!67.65^{+1.28}_{-1.26}\!\!$ & {$\!\!$flat $\beta=2\!\!$} & $\!\!1087.84/1116\!\!$ &$\!\!1095.84\!\!$&  {$\!\!67.81^{+1.12}_{-1.06}\!\!$}\rule{0pt}{1.4em} \\
\hline
 CPL   & all & $\!\!1084.28\!\!$ &$\!\!1094.28\!\!$& $\!\!67.76^{+1.15}_{-1.01}\!\!$ &  {$\!\!$flat CPL$\!\!$} & $\!\!1087.56/1116\!\!$ &$\!\!1095.56\!\!$&  {$\!\!67.99^{+1.11}_{-1.16}\!\!$}\rule{0pt}{1.4em} \\
\hline
 $\!\!w$CDM$\!\!$& all & $\!\!1085.19\!\!$ &$\!\!1093.19\!\!$ & $\!\!67.98^{+1.11}_{-1.11}\!\!$ & $\!\!$flat $w$CDM$\!\!$ & $\!\!1089.90/1117\!\!$ &$\!\!1095.90\!\!$&$\!\!68.16^{+1.12}_{-1.14}\!\!$ \rule{0pt}{1.4em} \\
\hline
$\!\!\Lambda$CDM$\!\!$  & all & $\!\!1089.42\!\!$ &$\!\!1095.42\!\!$& $\!\!69.02^{+1.04}_{-1.06}\!\!$ & {$\!\!$flat $\Lambda$CDM$\!\!$} &  $\!\!1089.96/1118\!\!$ &$\!\!1093.96\!\!$& $\!\!68.35^{+0.56}_{-0.55}\!\!$\rule{0pt}{1.4em}  \\
\hline
\end{tabular}
\end{table}

%


\section{Discussion}
\label{Disc}

We considered different cosmological models with variable equations of state (EoS) for
dark energy of the type  (\ref{war}) $p_x=w(a)\,\rho_x$, more precisely, models with EoS
(\ref{EoSn}),
 $$
w(a)= w_0+ w_1\frac{1-a^\beta}\beta a^\gamma,
 $$
generalizing the $\Lambda$CDM, $w$CDM,  Chevallier--Polarski--Linder (CPL), and
Barboza--Alcaniz--Zhu--Silva (BAZS) models~\cite{ChP,Linder,BarbozaAZS09}. These scenarios with
nonzero spatial curvature and with $\Omega_k=0$ were confronted with observational data
described in Section~\ref{Observ} and including SNe Ia, CMB data, and two classes of the
Hubble parameter estimates $H(z)$: from cosmic chronometers (CC) and from line-of-sight
baryonic acoustic oscillations ($H_\mathrm{BAO}$) data.

The results of our calculations for different models  are presented in
Tables~\ref{Estim} and \ref{EstAIC}, including minima of $\chi^2_\mathrm{tot}$, Akaike
information criterion (\ref{Akaike}), and the best fitted values with $1\sigma$
estimates of model parameters. We also investigated the linear model (\ref{linear}),
(\ref{Omlin}) with $N_p=5$ parameters (\ref{5param}), however it appeared to be
unsuccessful because it achieved the best fitted value of $\chi^2_\mathrm{tot}$ at $w_1$
very close to zero (see Figure~\ref{F1lin}). In~other words, the~linear model
(\ref{Omlin}) with the considered observational data was reduced to the $w$CDM model
with only $N_p=4$ parameters, but~both models have the same $\min\chi^2_\mathrm{tot}$.

Unlike the linear model (\ref{Omlin}) CPL scenario (\ref{CPL}), (\ref{OmCPL})  with the
same $N_p=5$ parameters (\ref{5param}) appeared to be more successful, in~particular,
for all $H(z)$ data, the CPL model yields $\min\chi^2_\mathrm{tot}\simeq1084.28$ in
comparison with $1085.19$ for the $w$CDM and $1089.42$ for the $\Lambda$CDM models. In~this case, the best fitted value $w_1\simeq0.62$ is far from zero, so CPL is not reduced
to the $w$CDM~model.

We should remember that a  large number $N_p$ of free model parameters is a drawback of
any model, and~when we use, in Table~\ref{EstAIC}, the Akaike information criterion
(\ref{Akaike}),  the $w$CDM  model with $\mathrm{AIC} = 1093.19$ will have the
advantage over CPL with $\mathrm{AIC} = 1094.28$ (and more essential advantage over the
$\Lambda$CDM model with $1095.42$). If~we consider the generalized model (\ref{EoSn}),
(\ref{OmEoSn}) with $N_p=7$ and additional parameters $\beta$ and $\gamma$, the~Akaike
expression (\ref{Akaike}) becomes too large and the model looks worse in comparison with
others.

\begin{figure}[ht]
 \hspace{-1.2cm}
\includegraphics[width=15.0 cm]{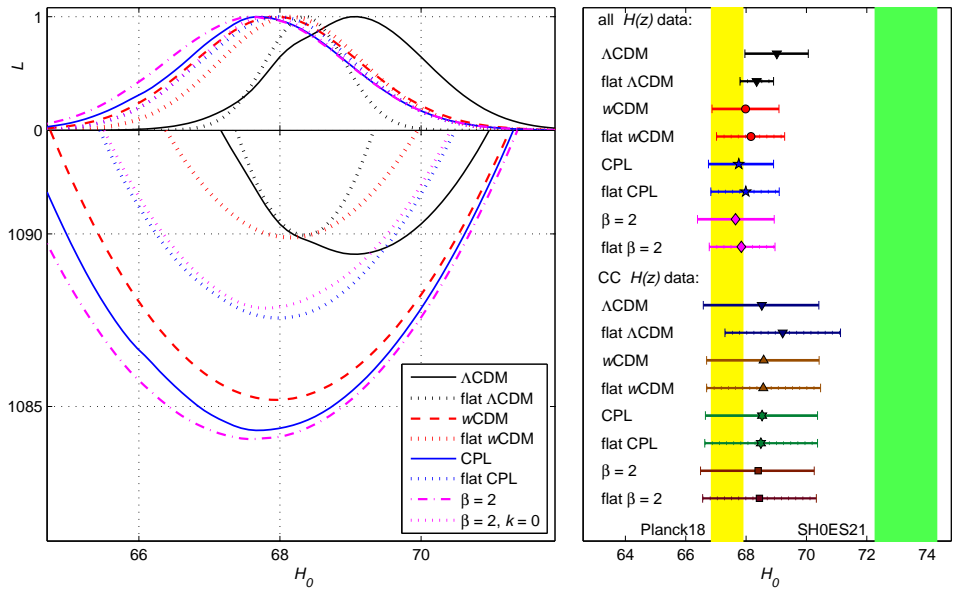}\vspace{-9pt}
\caption{One-parameter distributions $\chi^2_\mathrm{tot}(H_0)$ and likelihood functions
for models with all $H(z)$ data (the flat cases $k=0$ are shown with dotted lines);
$H_0$ estimates in the right panel are drawn as whisker plots.}
 \label{F5H}
\end{figure}

   {%
However, our analysis and Figure~\ref{F3gen} showed that the minimum
$\min\chi^2_\mathrm{tot}\simeq 1084.03$ of the generalized model (\ref{OmEoSn}) may be
achieved if we fix $\beta=2$, $\gamma=0$; the~resulting model ``$\beta=2$'' (\ref{be2})
has the same $N_p=5$ parameters (\ref{5param}) and absolutely minimal $\min\chi^2_\mathrm{tot}$
for all $H(z)$ data (its $\mathrm{AIC} =  1094.03$ is behind only the $w$CDM  AIC).
 }
The behavior of one-parameter distributions $\chi^2_\mathrm{tot}(H_0)$ and their minima for
all models with all $H(z)$ data are shown in  Figure~\ref{F5H}.

As mentioned above, if~we use only CC  $H(z)$ data, we observe smaller differences
between minima of $\chi^2_\mathrm{tot}$ for the considered models in Table~\ref{Estim}.
In this case, the Akaike criterion (\ref{Akaike}) gives advantage to the $\Lambda$CDM  model with
minimal $N_p$.

Table~\ref{Estim} and Figures~\ref{F2CPL} and \ref{F4be2} demonstrate that the success of
 the CPL scenario is achieved at  lower (best fitted) values of $\Omega_m^0$ and at larger values
of $\Omega_k$, if~we compare these results with the $w$CDM model. This tendency is
strengthened for the $\beta=2$  model (\ref{be2}). It looks natural, because~the CPL model
(\ref{OmCPL}) is the particular case of the BAZS model (\ref{OmBAZS}) with $\beta=1$.

One can see in Tables~\ref{Estim} and \ref{EstAIC} and Figures~\ref{F2CPL}, \ref{F4be2}, and
\ref{F5H} that the best fitted values of the Hubble parameter $H_0$ are very close for
models  $w$CDM, CPL, and $\beta=2$ in general and spatially flat cases. However, the~predictions of these models diverge with that of the $\Lambda$CDM model for all $H(z)$
data. The~last result is illustrated in Figure~\ref{F5H}; it  becomes more clear if we
look at Figure~\ref{F2CPL}, keeping in mind that  the $\Lambda$CDM model is the
particular case of  the $w$CDM model when $w=-1$. One should add the observed difference
between $H_0$ estimates of all models when we compare CC and all $H(z)$ data.

In the left panels of Figure~\ref{F5H} we draw one-parameter distributions
$\chi^2_\mathrm{tot}(H_0)$ and likelihood functions (\ref{likeli}) ${\cal L}(H_0)$ for
the considered models with all $H(z)$ data to clarify their best fits of $H_0$. For~the
considered four models we demonstrate here the results for the flat cases $k=0$
($\Omega_k=0$) as dotted~lines.

One can see that in the flat case $k=0$, the mentioned models yield appreciably larger
minima $\min\chi^2_\mathrm{tot}$, but~the best fitted values of $H_0$ change
unessentially for the $w$CDM, CPL, and $\beta=2$ models. For~the flat $\Lambda$CDM model,
the graph $\chi^2_\mathrm{tot}(H_0)$ is inscribed between the correspondent graphs of
the general $\Lambda$CDM and flat $w$CDM models; hence, the $H_0$ estimate for the flat
$\Lambda$CDM model lies between two estimates of the mentioned~scenarios.

All $H_0$ predictions of the $w$CDM, CPL, and $\beta=2$ models with all  $H(z)$ data fit
the Planck 2018 estimation of the Hubble constant~\cite{Planck18} (see
Figure~\ref{F5H}), but~they are far from the SH0ES 2021 estimation~\cite{Riess2021}. All
$H_0$ predictions with CC $H(z)$ data are very close for the four considered models and their
flat variants: they are larger but overlap the Planck 2018 value; however, they cannot
describe the tension with the SH0ES~data.

    {%
We can see from  Table~\ref{EstAIC} that the best values of 
AIC (\ref{Akaike}) for all $H(z)$ data demonstrate the $w$CDM and $\beta=2$
(\ref{be2}) models, but~for CC $H(z)$ data, the best  AIC has the flat $\Lambda$CDM model.
 }

\acknowledgments{G.S.S. is grateful to Sergei D. Odintsov for useful discussions and~support.}
\end{document}